\documentclass[12pt]{article}
\usepackage{hyperref}
\usepackage{amsmath}
\usepackage{amssymb}
\usepackage{subfig}
\usepackage{graphicx}
\usepackage{authblk}
\usepackage[symbol]{footmisc}
\usepackage{geometry}
\begin{document}

\begin{center}
\textbf{\large{Overspinning problem in Kerr black holes: second order corrections and self-energy}}
\medskip

\small{Iván R. Vásquez\footnote[2]{itsivanvasquez@gmail.com}} 
\medskip

\small{\textit{Grupo de Partículas Elementales y Cosmología, Departamento de Física, Universidad del Atlántico. Carrera 30 No. 8-49. Puerto Colombia, Atlántico, Colombia.}\footnote{This is the address corresponding to the time when the work \textit{was carried out}.}}

\end{center}

\begin{abstract}
We consider gedanken experiments to destroy Kerr black holes, by means of absorbing matter with sufficient energy and angular momentum. It is shown that extremal and near-extremal Kerr black holes cannot be destroyed in a process that includes a second order contribution to its final mass, and matter sources satisfy the null energy condition. Such contribution is calculated using hypersurface integration on the event horizon, and it traces similarities with terms related to matter-black hole interactions and a rotational self-energy lower bound suggested in previous works.
\end{abstract}
\vspace{1pc}

\section{Introduction}

Singularities are inevitably formed under certain conditions of gravitational collapse in general relativity, the proof came from Hawking \& Penrose with the celebrated \textit{singularity theorems} \cite{Penrose, Hawkingpenrose}. These cannot prove whether spacetime singularities are always hidden or exposed to observers at infinity. In response, Penrose suggested that an event horizon must hide the singularity so is never exposed to infinity, this is called the \textit{weak cosmic censorship conjecture} \cite{Penroseccc}. 

An attempt to test this conjecture would be possible if black holes can be destroyed (i.e. produce a naked singularity) by absorbing matter. Such an attempt was devised by Wald \cite{wald2}, assuming an absorption of test particles on a extremal Kerr-Newmann black hole. According to Wald's analysis, particles with too much charge or angular momentum are repelled either by electrostatic repulsion or spin-spin interaction, therefore preserving the cosmic censorship. Subsequent discussions found that nearly extremal black holes can absorb particles with definite values of charge and/or angular momentum, and become naked singularities \cite{hubeny, jacobson}. Such discussions make use of extended bodies as well, which tend to suffer very little perturbations coming from its own gravitational field and any other self-interactions. These works and others that followed, have concluded partially that these interactions may prevent black holes from being destroyed by disrupting any threatening process \cite{barau, zimmerman, Gao, Colleoni}. Collected, all these effects are commonly called \textit{backreaction effects}.

In an illuminating paper, Hod proposed that the amount of self-energy required to keep cosmic censorship valid, has a lower bound given by $L^2/8M^3$. Where $L$ is the absorbed angular momentum and $M$ the mass of the black hole \cite{hod}. More recently, Wald \& Sorce showed that these effects can be included into the process by calculating absorbed quantities at appropriate order \cite{waldsorce}. In particular, by expanding $M(\lambda)$ in terms of a real parameter $\lambda<1$ as $M(0)+\lambda\delta M+\lambda^2\delta^2 M+...$, one may include higher order terms of the pair $(\delta M, \delta J)$ contributing to the final mass of the black hole. The terms $\delta^{n}M$ corresponds to the $n^{th}$ perturbation on the initial mass, and they are better understood as integrals over $\mathcal{H}$.

In this paper, we consider a general description of absorption processes in near-extremal Kerr black holes. This description has been partially inspired by the work of Wald \& Sorce, and ultimately helps us to set up an overspinning scenario where the black hole is set to be destroyed. Our analysis allows us to confirm previous results and how higher order perturbations, such as $\delta^2 M$, prevent the black hole from being destroyed. Finally by exploring analogies with Reissner-Nordström black holes, a connection between $\delta^2M$, Hod's lower bound and other effects is found. Therefore, $\delta^2 M$ may contain all the information on backreaction effects as pointed out by Wald \& Sorce.

\section{Absorption processes in Kerr black holes}

In this section we establish a recipe to analyze absorption processes in Kerr black holes by employing a similar method as those in \cite{waldsorce, Natario}. We describe the process and consider every possible outcome.

Let us establish a simple description of the process:       (i) Initially, consider a Kerr family of solutions having a black hole region $\mathcal{B}$ parametrized by the pair $(M, J)$ satisfying the relation
\begin{align}\label{ineq}
M^2\geq J.
\end{align}
The Kerr family is the unique family of solutions to vacuum Einstein equations in stationary, asymptotically flat spacetime, parametrized by mass $M$ and angular momentum $J$. Whenever (\ref{ineq}) holds, the black hole region $\mathcal{B}$ is properly defined and forbids causal curves from reaching $\mathcal{I}^{+}$.

\begin{figure}[h!]
\centering
\includegraphics[scale=0.25]{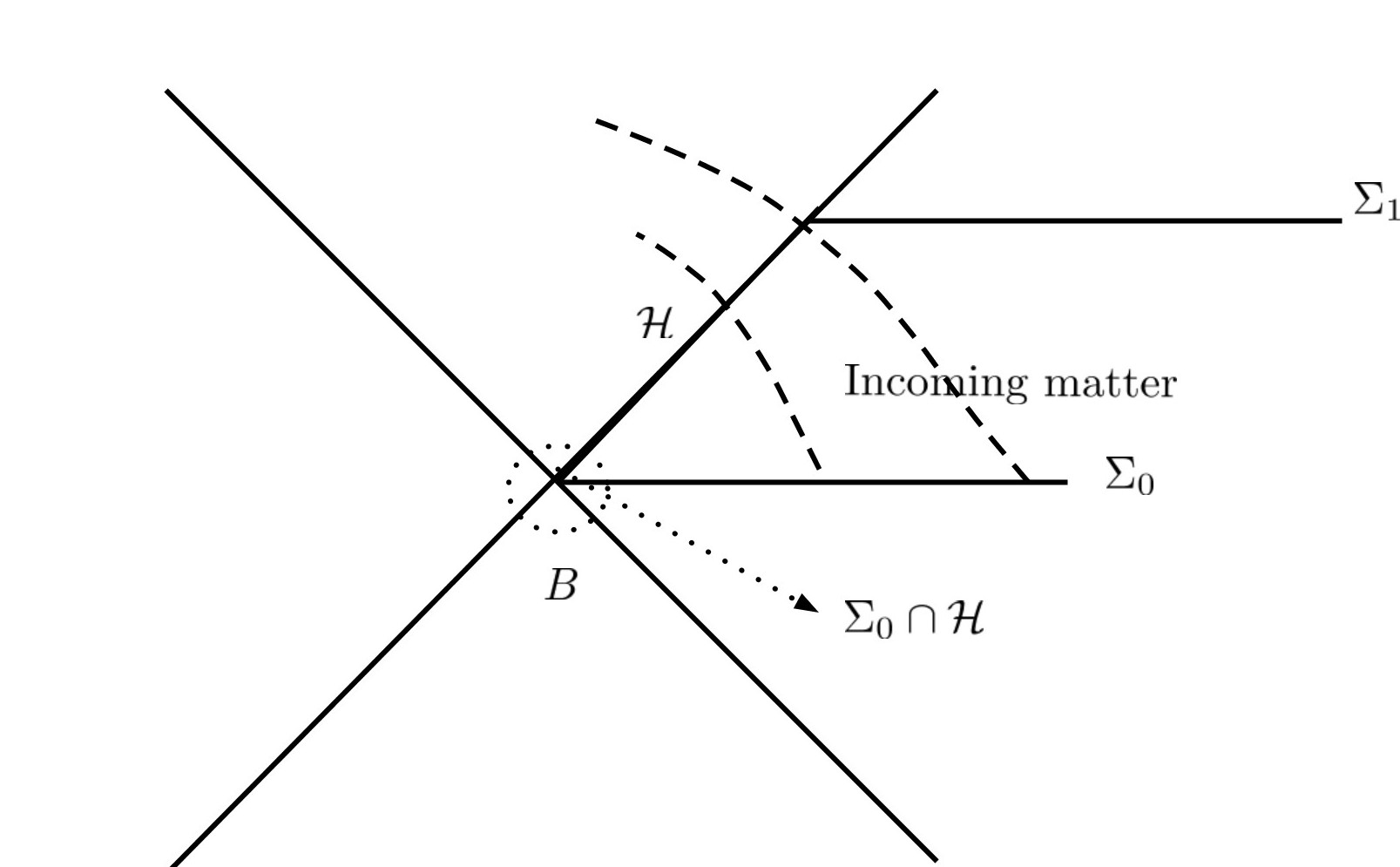} 
\caption{Conformal diagram of an absorption process in non-extremal Kerr black holes.}
\label{nonextremaloverspin}
\end{figure}

Now, (ii) spacetime is foliated with two spacelike hypersurfaces $\Sigma_{0}, \Sigma_{1}$, each one representing an instant of time during the process. Both are asymptotically flat and $\Sigma_{0}$ extends from the bifurcation surface $B$ to spatial infinity (see Figure \ref{nonextremaloverspin}). Accordingly, a quantity of matter comes near the horizon attempting to cross it, described by a \textit{test} energy-momentum tensor $T_{\mu\nu}$.

By following Wald \& Sorce, (iii) at early advanced times such matter causes a negligible perturbation on the black hole, meaning that in a neighbourhood of $\Sigma_{0}\cap \mathcal{H}$ the solution remains stationary. This makes possible to define the fluxes of energy and angular momentum density in terms of Killing vectors $t_{\mu}, \phi_{\mu}$ of the Kerr solution \cite{toolkit}:

\begin{align}
\varepsilon_{\mu}=-T_{\mu\nu}t^{\nu}, \quad l_{\mu}=T_{\mu\nu}\phi^{\nu}.
\end{align}
Since $T_{\mu\nu}$ is divergenceless and using Killing equations, both vector fields satisfy:

\begin{align}\label{gauss}
\oint_{\partial\mathcal{V}}\varepsilon_{\mu}d\Sigma^{\mu}=\oint_{\partial\mathcal{V}}l_{\mu}d\Sigma^{\mu}=0,
\end{align}
where integration is performed over a hypersurface $\partial\mathcal{V}$ enclosing the 4-volume which matter crosses in spacetime, thus we employ Gauss-Stokes theorem. As it is shown in Figure \ref{gaussintegral}, hypersurface $\partial\mathcal{V}$ is partitioned into several boundaries, and there exists a timelike hypersurface $S$ at which matter sources vanish. A sign convention is used to maintain every normal vector future-directed, giving the following:

\begin{figure}[h]
\centering
\includegraphics[scale=0.16]{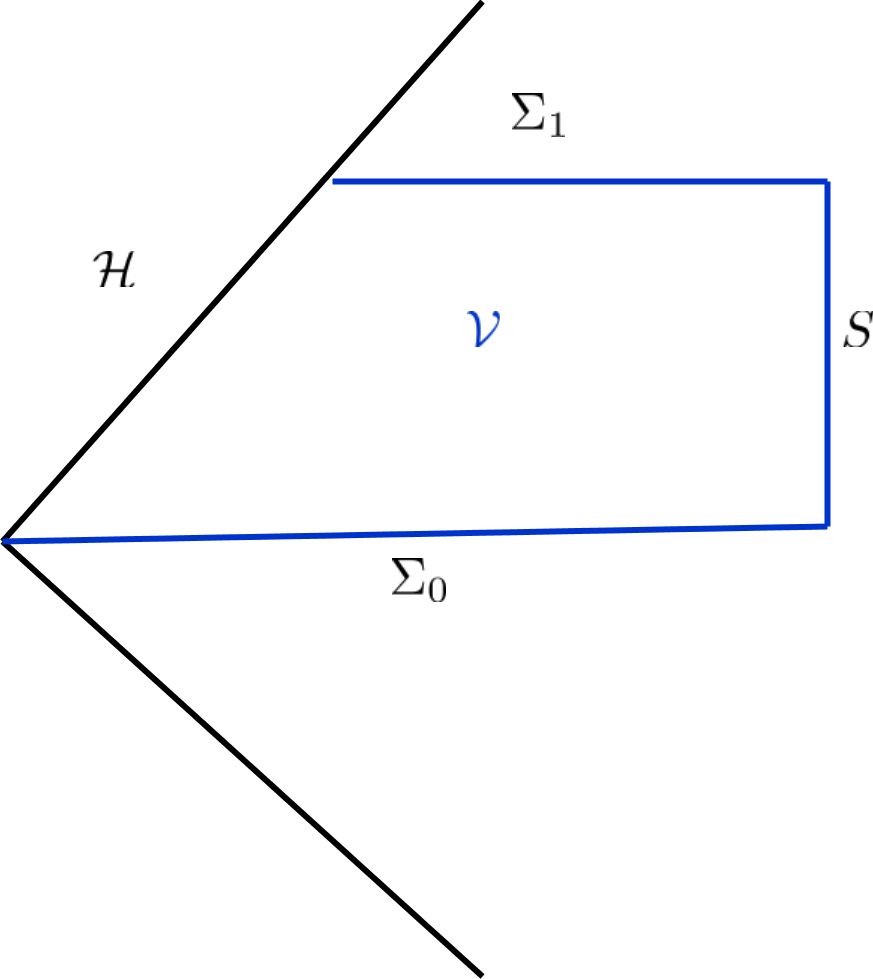} 
\caption{Integration volume and its boundaries. $S$ is a timelike hypersurface at infinity where all matter sources are assumed to vanish.}
\label{gaussintegral}
\end{figure}

\begin{align}
\int_{\Sigma_{1}}\varepsilon_{\mu}n^{\mu}_{1}\sqrt{h}d^3y-\int_{\Sigma_{0}}\varepsilon_{\mu}n^{\mu}_{0}\sqrt{h}d^3y=\int_{\mathcal{H}}\varepsilon_{\mu}d\Sigma^{\mu}, \quad \varepsilon_{\mu}\leftrightarrow l_{\mu}.
\end{align}
This shows that between $\Sigma_{0}, \Sigma_{1}$, a quantity of matter crosses $\mathcal{H}$ transferring energy and angular momentum to the black hole. This ultimately helps to define the infinitesimal quantities of mass ($\delta M$) and angular momentum $(\delta J)$ transferred through $\mathcal{H}$,

\begin{align}
-\int_{\mathcal{H}}d\Sigma_{\mu}T^{\mu}_{\nu}t^{\nu}=\delta M, \quad
\int_{\mathcal{H}}d\Sigma_{\mu}T^{\mu}_{\nu}\phi^{\nu}=\delta J.
\end{align}
Integration is performed over the whole event horizon $\mathcal{H}$, so that $d\Sigma_{\mu}=-dSdv\xi_{\mu}$, and by recalling stationary Killing vector $\xi_{\mu}:=t_{\mu}+\Omega_{H}\phi_{\mu}$, we combine $(\delta M, \delta J)$ giving the following relation:

\begin{align}\label{firstcross}
\delta M-\Omega_{H}\delta J=\int_{\mathcal{H}}d\Sigma T_{\mu\nu}\xi^{\mu}\xi^{\nu}; \quad d\Sigma:=d\Sigma_{\mu}\xi^{\mu}
\end{align}
In an ideal case, all the matter falls into the black hole contributing exactly to its mass and angular momentum and even if some of the matter remains in orbit. To us, it only matters the amount that crosses the event horizon. Given the contributions $(\delta M, \delta J)$ to the initial parameters $(M, J)$ we write:

\begin{align*}
M(\lambda)=M+\lambda\delta M+..., \\
J(\lambda)=J+\lambda\delta J+...
\end{align*}
In this sense, the pair $(M(\lambda), J(\lambda))$ represents a one-parameter family of Kerr solutions, where $\lambda\geq 0$ is the parameter whose values return a member of the family. Physically, can be interpreted as the fraction of matter absorbed at the end of the process since $\lambda=0$ identically returns the pair $(M, J)$ obeying (\ref{ineq}) at $\Sigma_{0}$. Note that also $\lambda\leq 1$, since $\lambda=1$ returns the case where quantities $(\delta M, \delta J)$ are completely absorbed by the black hole. Finally, let us define the following function in a similar fashion as in \cite{waldsorce}:

\begin{align}\label{firstf}
f(\lambda)=M^2-J+\lambda(2M\delta M-\delta J)+...,
\end{align}
note that higher order terms are neglected, allowing to use the pair $(\delta M,\delta J)$ whose relationship is known. In this case, $f(\lambda)$ gives a characterization for the family of Kerr solutions at $\Sigma_{1}$. At $\lambda=0$ then $f(0)>0$ represents the initial black hole solution at $\Sigma_{0}$, considering this, violations to (\ref{ineq}) will occur if:

\begin{align}
f(\lambda)<0; \quad \lambda\geq 0,
\end{align}
this is, whenever the final family of Kerr solutions is a naked singularity (overspinning occurs). It must be noted that the use of such function was introduced in Sorce-Wald \textit{gedanken experiments}, and subsequently used in other works to make similar conclusions \cite{Chen:2019nhv}. Cosmic censorship is preserved (the family contains black holes) if $f(\lambda)$ satisfies,

\begin{align*}
f(\lambda)\geq 0; \quad \lambda\geq 0.
\end{align*}
We can strictly forbid $f(\lambda)=0$ at $\Sigma_{1}$ by obeying the third law of black hole mechanics, but this is not an issue for now, so we relax the assumption.

\section{Extremal and non-extremal Kerr black hole: first order experiments}

In this section, the function $f(\lambda)$ in (\ref{firstf}) will contain only first order in $\lambda$, such that $\lambda<<1$ represents that a very small amount of matter is finally absorbed at $\Sigma_{1}$. This minimizes backreaction effects appearing near the horizon that may alter the process.
\\

\begin{figure}[h]
\centering
\includegraphics[scale=0.18]{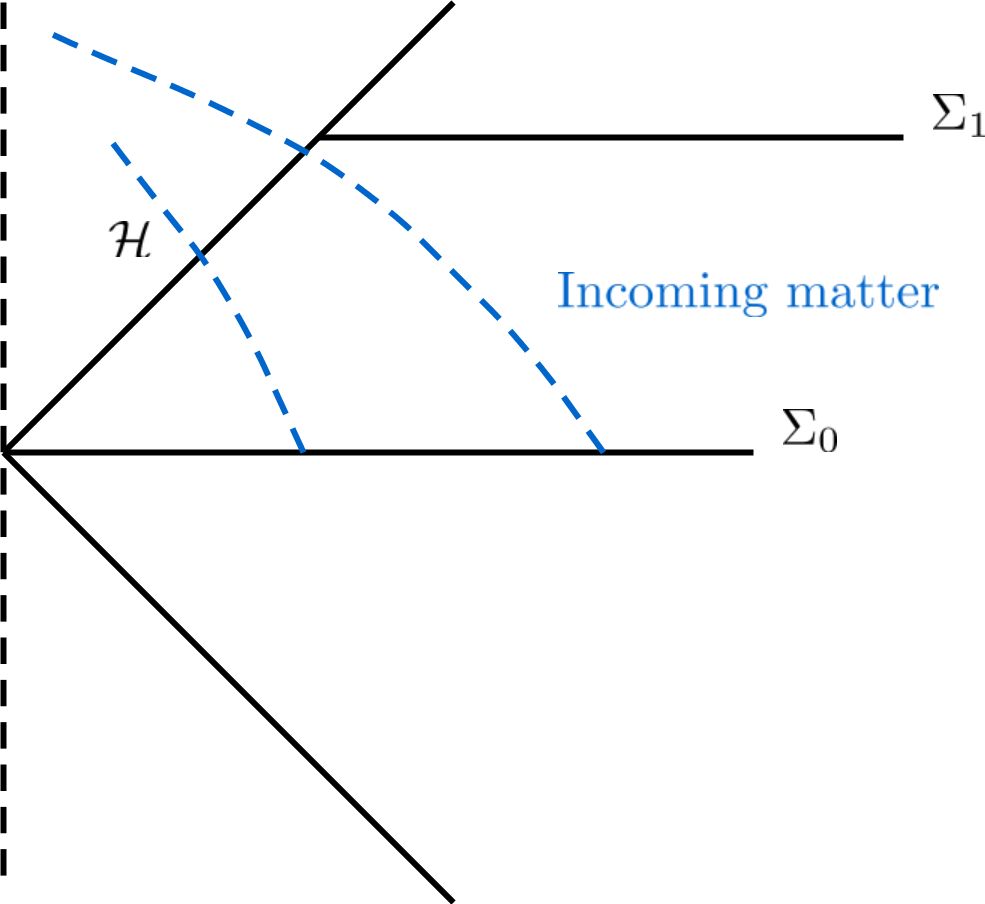} 
\caption{Matter absorption in a extremal black hole. $\Sigma_{0}$ extends from the horizon to spatial infinity, in a neighbourhood of $\Sigma_{0}\cap \mathcal{H}$ perturbations are negligible.}
\label{extremaloverspin}
\end{figure}

Now, a saturation in (\ref{ineq}) represents extremal black hole solutions, which give a maximum rate of angular momentum for black holes to exist (i.e. $J_{max}=M^2)$. In such case, we have a conformal diagram as follows (Figure \ref{extremaloverspin})

In this case our initial condition at $\Sigma_{0}$ is $f(0)=0$ from (\ref{firstf}), where $f(\lambda)$ takes the simple form:

\begin{align}
f(\lambda)=\lambda(2M\delta M-\delta J).
\end{align}
Similarly for a extremal black hole, the relation (\ref{firstcross}) reduces to:

\begin{align}\label{firstcrossex}
\delta M-\frac{\delta J}{2M}=\int_{\mathcal{H}}T_{\mu\nu}\xi^{\mu}\xi^{\nu}d\Sigma,
\end{align}
where the factor $1/2M$ comes from the angular velocity $\Omega_{H}$ with $J=M^2$. By knowing that any violation to (\ref{ineq}) occurs whenever $f(\lambda)<0$, we conclude that extremal Kerr black holes cannot be destroyed if $T_{\mu\nu}$ satisfies the null energy condition $T_{\mu\nu}\xi^{\mu}\xi^{\nu}\geq 0$, thus (\ref{firstcrossex}) is positive. Such result has been recognized widely in literature \cite{wald2, hubeny, jacobson, Gao, waldsorce, Natario}.

Now, we turn our attention to non-extremal black holes obeying (\ref{ineq}) in a restricted form ($M^2>J$). In particular, we treat the case for a black hole in the verge of becoming extremal (i.e. \textit{near extremal}), so (\ref{ineq}) can be expressed as:

\begin{align}
J\approx M^2(1-2\epsilon^2),
\end{align}
here $\epsilon$ is a positive scalar factor, such that $\epsilon<<1$. This choice has been inspired by its first use in \cite{hubeny} and later in \cite{jacobson}. It also simplifies many relevant quantities, such as the outer horizon radius $r_{+}=M(1+2\epsilon)$. The appropriate diagram for the process is figure \ref{nonextremaloverspin}, with initial condition $f(0)=2\epsilon^2M^2>0$ at $\Sigma_{0}$, and $\Omega_{H}=(1-2\epsilon-2\epsilon^2)/2M+O(\epsilon^3)$. Therefore relation (\ref{firstcross}) becomes,

\begin{align}
\delta M-\frac{\delta J}{2M}-\frac{\epsilon+\epsilon^2}{2M}=\int_{\mathcal{H}}d\Sigma T_{\mu\nu}\xi^{\mu}\xi^{\nu}.
\end{align}
When $T_{\mu\nu}$ satisfies the null energy condition, $f(\lambda)$ acquires a lower bound whose expression reads,

\begin{align}
f(\lambda)\geq 2\epsilon^2M^2-2\lambda\epsilon\delta J+O(\lambda\epsilon^3).
\end{align}
From this, is noted that parameters $\epsilon, \lambda, \delta J$ can be adjusted for overspinning to take place, allowing $f(\lambda)$ to attain negative values by making the RHS negative. Accordingly, overspinning will occur if $\lambda\delta J>M^2\epsilon$, so we take $\epsilon\sim \alpha\lambda\delta J/M^2$ for $\alpha<1$, note that this condition is only valid for $\epsilon>0$ (i.e. non-extremal black holes). Hence, a violation to (\ref{ineq}) is found to correspond to a second order in $\lambda$, that is,

\begin{align}
f(\lambda)\gtrsim 2\lambda^2\frac{(\delta J)^2}{M^2}\alpha(\alpha-1).
\end{align}
Since $\alpha<1$, it is assured that the RHS is negative. However, the resulting violation is of second order in $\lambda$, suggesting the need to include $\lambda^2$ terms in the final pair $(M(\lambda), J(\lambda))$. As this is not included in our first description of the process, we see that this result hints a modification in the process to fully describe the resulting family of solutions at $\Sigma_{1}$. In this sense, as it has been shown by Wald \& Sorce \cite{waldsorce}, any \textit{gedanken experiment} devised to destroy black holes cannot be complete if second order terms are neglected.

\section{Absorption processes in Kerr black holes with second order corrections}

In order to include second order corrections, the process is more or less the same, however some changes are introduced for convenience.

\begin{figure}[h]
\centering
\includegraphics[scale=0.16]{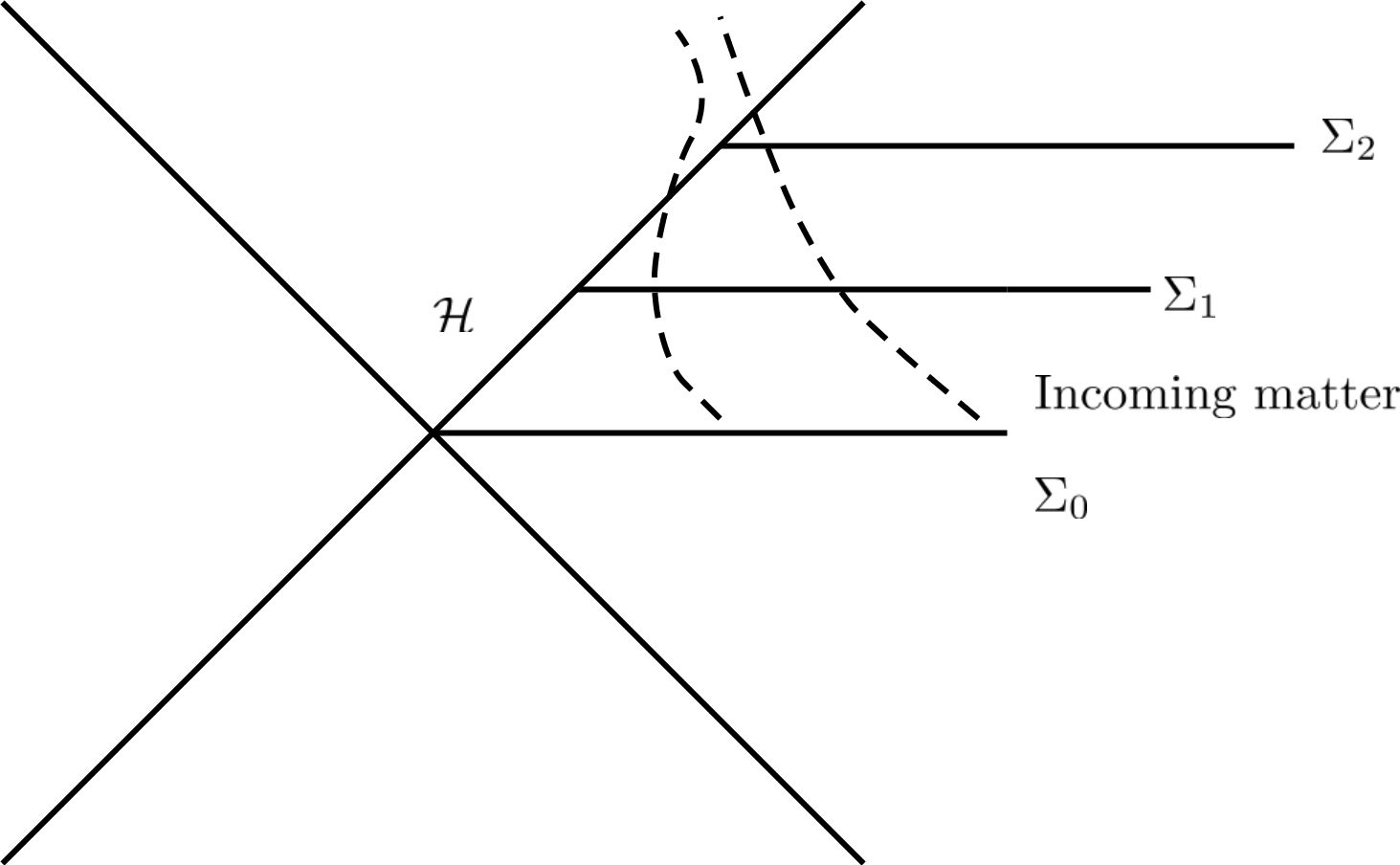} 
\caption{Conformal diagram for an absorption process in non-extremal black hole. Foliated with spacelike hypersurfaces $\Sigma_{0}, \Sigma_{1}, \Sigma_{2}$ with an adiabatic absorption at a neighbourhood of cross-section $\Sigma_{1}\cap\mathcal{H}$}
\label{overspinsecond}
\end{figure}

Let us establish the following modifications: (i) Matter evolution is followed by defining a hypersurface $\Sigma$ parallel to $\mathcal{H}$, from the cross-section $\Sigma_{0}\cap \mathcal{H}$ to cross-section $\Sigma_{1}\cap\mathcal{H}$ where matter has fallen slowly and \textit{adiabatically} to the horizon (see Figure \ref{overspinsecond}).

Accordingly, the term \textit{adiabatic} is used in the sense of a process for which $\delta A_{H}=0$, which in turn is \textit{reversible} according to Christodoulou \cite{Christodoulou}. Therefore, recalling the first law of black hole dynamics \cite{bch}, we get:

\begin{align}\label{adiabatic}
\delta M=\Omega_{H}\delta J.
\end{align}
Thus, according to (\ref{firstcross}) matter is brought up to $\mathcal{H}$ without being absorbed. Recalling test particle motion, the expression for $\dot{r}$ in the equatorial plane vanishes at $\mathcal{H}$ for the Kerr metric whenever $E=\Omega_{H}L$. In the spirit of this result, we may think that infalling matter reaches a `contact point' with the event horizon, and it's at this `point' where a further interaction with the null generators of $\mathcal{H}$ may occur, this has been partially inspired by its use in \cite{hod}.

Now, (ii) since adiabatic condition implies $T_{\mu\nu}\xi^{\mu}\xi^{\nu}=0$ (there is no net flux of matter through $\mathcal{H}$), after reaching contact point the energy-momentum tensor must be modified with an additional contribution, namely $T_{\mu\nu}(\Sigma_{1}\cap\mathcal{H})\sim\beta\delta T_{\mu\nu}$. In correspondence with previous process, such additional contribution vanishes at early advanced times, that is, $T_{\mu\nu}(\Sigma_{0}\cap\mathcal{H})\sim T_{\mu\nu}(0)$. 

\begin{figure}[h]
\centering
\includegraphics[scale=0.16]{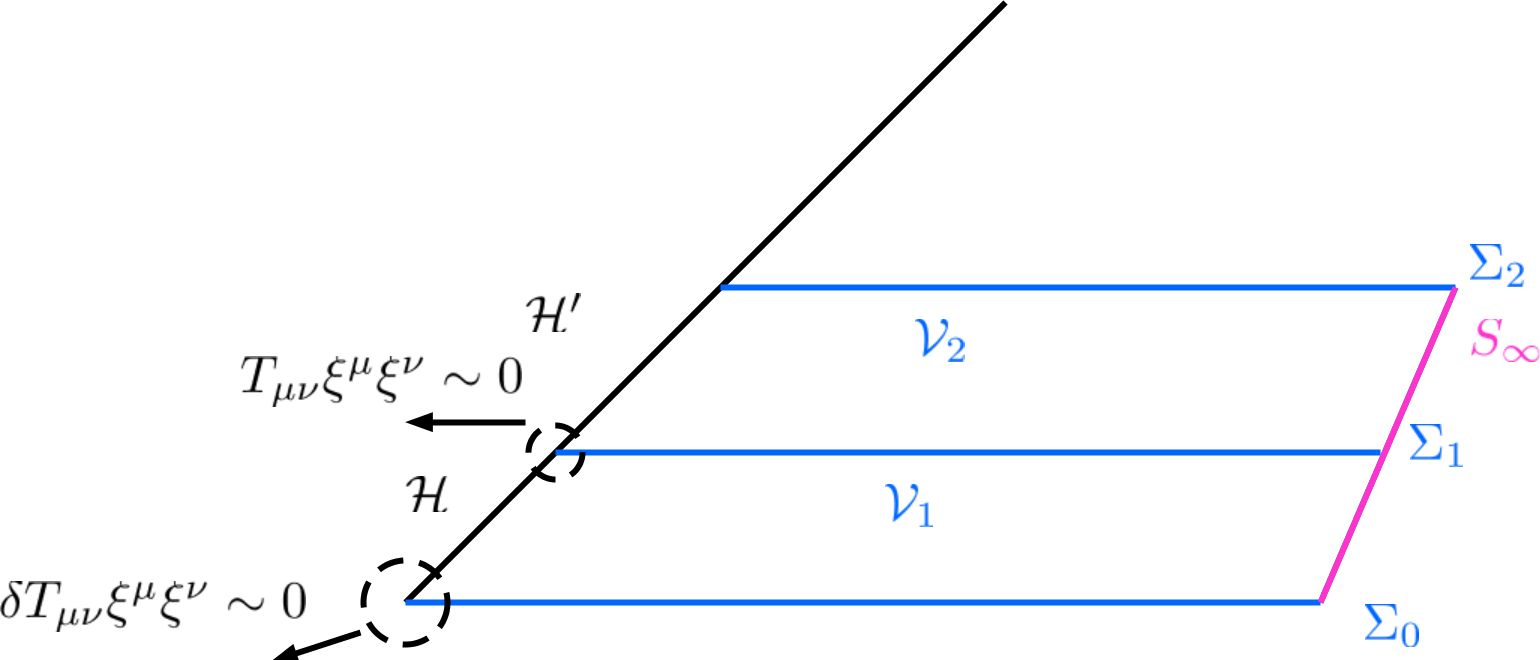} 
\caption{Integration volume and its boundaries. $S_{\infty}$ is a timelike hypersurface where all matter sources are assumed to vanish or possibly radiated if are not absorbed.}
\label{intvolume2}
\end{figure}

At this point we can define the pair $(\delta \varepsilon_{\mu}, \delta l_{\mu})$ and apply Gauss-Stokes theorem to the modified energy-momentum tensor $T_{\mu\nu}(\beta)$,

\begin{align}
\oint_{\mathcal{V}}\varepsilon_{\mu}(\beta)d\Sigma^{\mu}=\oint_{\mathcal{V}_{1}}\varepsilon_{\mu}(0)d\Sigma^{\mu}+\beta\oint_{\mathcal{V}_{2}}\delta\varepsilon_{\mu}d\Sigma^{\mu}=0, \quad \varepsilon_{\mu}\leftrightarrow l_{\mu}.
\end{align}
Integration volumes are shown in Figure \ref{intvolume2}, and similarly we can partition closed integrals into,

\begin{multline}
\int_{\mathcal{H}}\varepsilon_{\mu}d\Sigma^{\mu}+\beta\int_{\mathcal{H}'}\delta\varepsilon_{\mu}d\Sigma^{\mu}=\int_{\Sigma_{1}}\varepsilon_{\mu}n^{\mu}_{1}\sqrt{h}d^3 y-\int_{\Sigma_{0}}\varepsilon_{\mu}n^{\mu}_{0}\sqrt{h}d^3 y+\\\beta\left(\int_{\Sigma_{2}}\delta\varepsilon_{\mu}n^{\mu}_{2}\sqrt{h}d^3 y-\int_{\Sigma_{1}}\delta\varepsilon_{\mu}n^{\mu}_{1}\sqrt{h}d^3 y-\int_{S_{\infty}}\delta\varepsilon_{\mu}d\Sigma^{\mu}\right).
\end{multline}
Again, the same is true for the fields $l_{\mu}, \delta l_{\mu}$ and we similarly couple with the aid of the Killing field $\xi^{\mu}$:

\begin{multline}
\int_{\mathcal{H}}T_{\mu\nu}\xi^{\mu}\xi^{\nu}d\Sigma+\beta\int_{\mathcal{H}'}\delta T_{\mu\nu}\xi^{\mu}\xi^{\nu}d\Sigma'=\int_{\Sigma_1}\varepsilon_{\mu}n^{\mu}_{1}\sqrt{h}d^3y-\int_{\Sigma_0}\varepsilon_{\mu}n^{\mu}_{1}\sqrt{h}d^3y+ \\
\beta\left(\int_{\Sigma_2}\delta\varepsilon_{\mu}n^{\mu}_{2}\sqrt{h}d^3y-\int_{\Sigma_1}\delta\varepsilon_{\mu}n^{\mu}_{1}\sqrt{h}d^3y-\int_{S_{\infty}}\delta\varepsilon_{\mu} d\Sigma^{\mu}\right).
\end{multline}

The first term to the left corresponds to relation (\ref{firstcross}), which vanishes in virtue of the adiabatic condition (\ref{adiabatic}). The immediate implication results since $A_{H}$ is a constant and now $\delta T_{\mu\nu}\xi^{\mu}\xi^{\nu}$ is integrated all over $\mathcal{H}$, resulting in the following conditions:

\begin{align}
\label{conservationq}
\int_{\Sigma_1}T_{\mu\nu}\xi^{\nu}n^{\mu}_{1}\sqrt{h}d^3y=\int_{\Sigma_0}T_{\mu\nu}\xi^{\nu}n^{\mu}_{0}\sqrt{h}d^3y, \\
\label{secondflux}
\int_{\mathcal{H}}\delta T_{\mu\nu}\xi^{\mu}\xi^{\nu}d\Sigma=\int_{\Sigma_2}\delta T_{\mu\nu}\xi^{\nu}n^{\mu}_{2}\sqrt{h}d^3y-\int_{\Sigma_1}\delta T_{\mu\nu}\xi^{\nu}n^{\mu}_{1}\sqrt{h}d^3y-\int_{S_{\infty}}\delta T_{\mu\nu}\xi^{\nu}d\Sigma^{\mu}.
\end{align}
The condition (\ref{conservationq}) reflects the energy-momentum conservation in volume $\mathcal{V}_{1}$, since matter does not fall through $\mathcal{H}$. Since at a neighbourhood of $\Sigma_{1}\cap \mathcal{H}$ an interaction is assumed to occur, matter falls through the event horizon after that and its flux is given by (\ref{secondflux}). The last term in the RHS of (\ref{secondflux}) represents the flux of matter through a timelike hypersurface at infinity, in the case that matter is radiated after the interaction. 

Now we can impose the energy-momentum tensor $\delta T_{\mu\nu}$ to satisfy the null energy condition and obtain the simultaneous relations:

\begin{align}
\label{nullencond}
\int_{\mathcal{H}}\delta T_{\mu\nu}\xi^{\mu}\xi^{\nu}d\Sigma\geq 0, \\
\label{fluxdiff}
\int_{\Sigma_2}\delta T_{\mu\nu}\xi^{\nu}n^{\mu}_{2}\sqrt{h}d^3y-\int_{\Sigma_1}\delta T_{\mu\nu}\xi^{\nu}n^{\mu}_{1}\sqrt{h}d^3y\geq\int_{\infty}\delta T_{\mu\nu}\xi^{\nu}d\Sigma^{\mu}.
\end{align}
Hence, (\ref{fluxdiff}) gives the possibility that if $\delta T_{\mu\nu}\xi^{\mu}\xi^{\nu}=0$ there will be no net flux through $\mathcal{H}$ and all matter is radiated away to infinity. Since we are interested in the case that matter entirely crosses the horizon, we turn to the null energy condition (\ref{nullencond}) by proposing that such interaction gives rise to a change in the angular velocity $\delta\Omega_{H}$, which can be expressed in terms of a perturbation on the Killing vector field $\delta\xi^{\mu}$ while keeping temporal and axial symmetry (i.e. $\delta \xi^{\mu}=\delta \Omega_{H}\phi^{\mu}$)\footnote{This can be interpreted as a very small modification of the metric during the process, such that, $\delta t^{\mu}, \delta \phi^{\mu}\sim 0$ which hints at a very slow process without disrupting equilibrium. }. To use this, we rewrite:

\begin{align}
\int_{\mathcal{H}}\delta T_{\mu\nu}\xi^{\mu}\xi^{\nu}d\Sigma=\int_{\mathcal{H}}\delta T_{\mu\nu}\xi^{\mu}t^{\nu}d\Sigma+\Omega_{H}\int_{\mathcal{H}}\delta T_{\mu\nu}\xi^{\mu}\phi^{\nu}d\Sigma.
\end{align}
Now, we rewrite each term using product of variations and obtain:

\begin{align}
\delta\left(-\int_{\mathcal{H}}T_{\mu\nu}t^{\nu}d\Sigma^{\mu}\right)-\int_{\mathcal{H}}\delta T_{\mu\nu}t^{\nu}\delta \xi^{\mu}d\Sigma=\int_{\mathcal{H}}\delta T_{\mu\nu}t^{\mu}\xi^{\nu}d\Sigma =\delta^2 M-\delta \Omega_{H}\int_{\mathcal{H}}T_{\mu\nu}t^{\nu}\phi^{\mu}d\Sigma.
\end{align}
The same is true for angular momentum,

\begin{align}
-\delta\left(\int_{\mathcal{H}}T_{\mu\nu}\phi^{\nu}d\Sigma^{\mu}\right)-\int_{\mathcal{H}}\delta T_{\mu\nu}\phi^{\nu}\delta \xi^{\mu}d\Sigma=\int_{\mathcal{H}}\delta T_{\mu\nu}\phi^{\mu}\xi^{\nu}d\Sigma =-\delta^2 J-\delta \Omega_{H}\int_{\mathcal{H}}T_{\mu\nu}\phi^{\nu}\phi^{\mu}d\Sigma,
\end{align}
the quantities $\delta^2 M, \delta^2 J$ can interpreted as second order perturbations in the pair $(M, J)$ or variations of their fluxes across $\mathcal{H}$. Finally, plugging back into (\ref{nullencond}), an important relation is obtained:

\begin{align}\label{secondcrossit}
\delta^2 M-\Omega_{H}\delta^2J\geq \delta \Omega_{H}\delta J.
\end{align}
Here the interpretation is somewhat clear; $\delta^2 M$ arises in response to a change in the rotational energy $\Omega_{H}\delta J$ where the null generators of the horizon tend to rotate faster by an amount $\delta \Omega_{H}$ (i.e. $\Omega_{H}\rightarrow \Omega_{H}+\delta \Omega_{H}$). All of this is assumed to occur at a `contact point' between matter and the horizon when it falls adiabatically. This change can be though to arise by noting that $\delta^2 M \geq \delta(\Omega_{H}\delta J)$ exactly, and since a quantity of matter (or particle) crosses $\mathcal{H}$ whenever $\delta M\geq \Omega_{H}\delta J$ (the flux across $\mathcal{H}$ is positive), it is natural to think that if the rotational energy by the black hole is increased all matter sources must overcome a higher `energy barrier' to cross the horizon. Also the flux of the field vector $\varepsilon_{\mu}$ is affected, something that can only happen if very near $\mathcal{H}$ a contribution to energy arises from matter itself (no other interactions other than gravity are present), this is what we argue would be `self-energy'. 

Now, a variation of $\Omega_{H}$ will generally contain variations of the pair $(\delta M, \delta J)$ and in order to simplify we use the following form for $\Omega_{H}$:

\begin{align}
\Omega_{H}=\frac{4\pi J}{A_{H} M},
\end{align}
then, given the adiabatic condition there is no variation on the surface area $A_{H}$. The total variation $\delta \Omega_{H}$ is given by:

\begin{align}\label{will}
\delta \Omega_{H}=\frac{\delta J}{4M^3}\left(\frac{2M^3r_{+}-J^2}{M^2r^{2}_{+}}\right).
\end{align}
Where we've used adiabatic condition (\ref{adiabatic}) to eliminate $\delta M$ and now $\delta \Omega_{H}$ is directly proportional to $\delta J$, the angular momentum of matter. From this relation we can observe that $\delta\Omega_{H}\rightarrow \delta J/4M^3$ at a slowly-rotating regime ($J\rightarrow 0$) and the solution approaches Schwarzschild; therefore a black hole with zero angular momentum can form an ergoregion, acquiring non-zero angular velocity in the presence of rotating matter outside. Such an effect has been studied by Will, by considering stationary axisymmetric distributions of matter outside a slowly-rotating black hole and showed how a ring of matter can effectively induce an angular velocity of the same form as (\ref{will}) in such regime \cite{will, will2}.

Similarly, in the near-extremal regime we can take $J=M^2(1-2\epsilon^2)$ and $\delta \Omega_{H}$ can be expressed exactly as,

\begin{align}\label{changeomega}
\delta \Omega_{H}=\frac{\delta J}{4M^3}\left(1+\frac{\epsilon^4}{1+4\epsilon+4\epsilon^2}\right).
\end{align}
Here we can ignore higher orders of $\epsilon^2$ and essentially $\delta \Omega_{H}\approx \delta J/4M^3$, the same as in the slowly-rotating black hole case. This suggests that even in a rapidly-rotating black hole, the adiabatic \textit{assimilation} of matter produces an inertial frame-dragging effect, causing null generators to rotate faster, presumably pushing it towards extremality. In particular, induces a change $\delta \Omega_{H}$, which is independent of the black hole's angular momentum and only depends on the quantity $\delta J$, in a relationship noted to be \textit{universal} and obtained by analyzing a process where a ring of matter is adiabatically assimilated (or absorbed) by a central black hole in \cite{hod2}; such work predicts a smooth transition from the initial $\Omega_{H}$ to the final value $\Omega_{H}+\delta \Omega_{H}$.

Finally, we plug (\ref{changeomega}) in (\ref{secondcrossit}) to obtain,

\begin{align}\label{secondcrossitoff}
\delta^2 M-\Omega_{H}\delta^2 J \gtrsim \frac{(\delta J)^2}{4M^3}.
\end{align}
Therefore, second order perturbations on $(M, J)$ are effectively bounded below by a term quadratic in $\delta J$ representing an interaction term, as a consequence of the null energy condition. This second order correction can be added to the function $f(\lambda)$ defined previously in (\ref{firstf}) up to first order, whose expanded form up to $\lambda^2$ is:

\begin{align}
f(\lambda)=M^2-J+\lambda(2M\delta M-\delta J)+\lambda^2(2M\delta^2 M-\delta^2J+\delta M^2),
\end{align}
using adiabatic condition (\ref{adiabatic}) we can eliminate $\delta M$,

\begin{align}
f(\lambda)=M^2-J+\lambda(2M\Omega_{H}-1)\delta J+\lambda^2(2M\delta^2 M-\delta^2J+\Omega^2_{H}\delta J^2).
\end{align}
A lower bound is established by relation (\ref{secondcrossitoff}), which is merely a consequence of imposing the null energy condition on $\delta T_{\mu\nu}$, so $f(\lambda)$ acquires the following lower bound:

\begin{align}
f(\lambda)\gtrsim 2\epsilon^2M^2-2\lambda\epsilon\delta J+\lambda^2\frac{3(\delta J)^2}{4M^2}+O(\lambda^2\epsilon,\lambda^2\epsilon^2).
\end{align}
Since $\epsilon, \lambda<<1$ we can ignore mixed terms between them, and especially those of higher orders. Finally, the whole relation can be written as a perfect square plus a positive quantity,

\begin{align}
f(\lambda)\gtrsim 2\left(M\epsilon-\lambda\frac{\delta J}{2M}\right)^2+\lambda^2\frac{(\delta J)^2}{4M^2}+O(\lambda^2\epsilon, \lambda^2\epsilon^2),
\end{align}
which is evidently positive for any $\lambda\geq 0$ and no violation to (\ref{ineq}) occurs, in agreement with the weak cosmic censorship conjecture. From this result, it is worth noting that $f(\lambda)$ cannot pass through zero for any value of $\lambda$, meaning that extremal Kerr solutions cannot result from this process in agreement with the third law of black hole mechanics \cite{Duztas:2017ycz}.

\section{Discussion}
In this paper we showed that gedanken experiments to destroy Kerr black holes cannot accomplish their goal, by analyzing an absorption process where matter is adiabatically lowered to the horizon and brought down to a `contact point' where an interaction between infalling matter and the null generators occur. Such interaction results in a faster rotation or angular velocity $\Omega_{H}$ as it is suggested in \cite{hod}. An important aspect of this experiment is the assumption that the energy-momentum tensor can be written in the form $T_{\mu\nu}(\beta)=T_{\mu\nu}+\beta\delta T_{\mu\nu}$, where $\delta T_{\mu\nu}$ is the energy-momentum tensor after interaction, this is similar to the method employed in \cite{waldsorce} (the latter being more rigorously applied). In essence, infalling matter is absorbed fully after interaction, and previous results suggest that a \textit{smooth} transition between angular velocities $\Omega_{H} \rightarrow\Omega_{H}+\delta \Omega_{H}$ occur \cite{hod2} in the black hole-ring composite system. The term $\delta\Omega_{H}$ has a definite asymptotic value for near-extremal black holes given by $\delta J/4M^3$, and corresponds to an inertial dragging effect induced by matter on the central Kerr black hole. Now our final remark is centered in the right hand side of inequality (\ref{secondcrossitoff}) since we can apply the same methods above to evaluate a possible `overcharging' scenario in Reissner-Nostrdöm black holes. Virtually this can turn simple by exploiting certain dualities that both metrics share: $a\rightarrow Q$, $\delta J \rightarrow \delta Q$, $\Omega_{H}\rightarrow \Phi_{H}$. Doing this, results in a non-rotating charged version of (\ref{secondcrossitoff}):

\begin{align}\label{secondrn}
\delta^2 M-\Phi_{H}\delta Q \gtrsim \frac{(\delta Q)^2}{M}(1-2\epsilon)+O(\epsilon^2).
\end{align}
This result is verified when we take $\delta J=\delta^2J=J=0$ in the results of Wald \& Sorce, whose calculations are rigorously carried out \cite{waldsorce}. The behaviour of solutions at $\Sigma_{2}$ (the end of the process) is the same as in Kerr, it contain black holes. The remark shall be made after considering the following: (i) the energy required to bring a charge $\delta Q$ from infinity to the polar axis $\theta=0$ at $\mathcal{H}_{+}$ in a Reissner-Nordström has been calculated in \cite{lohiya}, given by:

\begin{align}
W=\frac{Q\delta Q}{r_{+}}+\frac{M(\delta Q)^2}{2r^2_{+}}.
\end{align}
The second term is the self-energy term due to corrections for \textit{polarization} of the horizon, which results in a repulsive force (out from the black hole) by an image charge induced inside the horizon (See \cite{hod, linet, Hod3}). In the near-extremal regime becomes $E_{sf}\approx(\delta Q)^2(1-4\epsilon)/2M$. Another contribution would be the redshifted contribution to the rest mass, in the case of a spherical distribution of charge \cite{waldsorce, hod}, it has a lower bound given by $E_{s}\geq \kappa_{+}(\delta Q)^2/2$ since the amount of work done $\delta W$ depends on a proper length interval $\delta s$ (i.e. $\delta W\sim\kappa E_{0}\delta s; \quad s\geq R$), where $E_{0}=(\delta Q)^2/2R$. In the near-extremal regime both can be added to give the RHS of (\ref{secondrn}):

\begin{align}
E_{s}+E_{SF}\gtrsim \frac{\epsilon(\delta Q)^2}{M}+\frac{(\delta Q)^2(1-4\epsilon)}{2M}+O(\epsilon^2)=\frac{(\delta Q)^2}{2M}(1-2\epsilon)+O(\epsilon^2),
\end{align}
where we used $\kappa_{+}\approx 2\epsilon(1-2\epsilon)/M$. As it is shown, the sum of contributions from self-energy and finite size result in half the expected value, therefore $2(E_{SF}+E_{s})=\delta^2 M$. Let us denote $\delta^2 M/2=E^{(2)}$ and take $\delta^2Q=0$ without loss of generality, we do it so the actual energy coming from matter has the following form:

\begin{align}
E=\delta M+\frac{1}{2}\delta^2 M+...
\end{align}
The factor of $1/2$ corresponds to the Taylor expansion. Finally, applying this result to the Kerr metric, we obtain that an energy $E^{(2)}$ with contributions such as self-energy, must have a lower bound given exactly by:

\begin{align}
E^{(2)}\geq \frac{(\delta J)^2}{8M^3}.
\end{align}
This lower bound was obtained in \cite{hod, needham} by assuming cosmic censorship and area theorem. Here we proved that arises naturally to prevent a black hole from violating weak cosmic censorship, provided that matter sources satisfy the null energy condition after interaction with the black hole, in this spirit it reinforces the idea that weak cosmic censorship prevails in many counterexamples explored \cite{Hod4}. Most recently, Wald and Mackewicz have rigorously proven such lower bound, for spinning bodies in the Kerr metric \cite{macke}. 

\section*{Acknowledgements}

The present paper are the results obtained for the undergraduate dissertation of I.R.V. under the supervision of Erick Tuirán (Universidad del Norte, Barranquilla). I.R.V would like to thank Universidad del Atlántico and the research group PEyCOS (Partículas elementales y Cosmología) to which I.R.V was affiliated during his time as an undergraduate student. I.R.V would like to thank Erick Tuirán, for fruitful discussions regarding several parts of this work.


\begin{thebibliography}{99}
\bibitem{Penrose} 
  R. Penrose, Phys.\ Rev.\ Lett. {\bf 14}, 57 (1965).
  
\bibitem{Hawkingpenrose} 
  S.W Hawking and R. Penrose, Proc.\ R.\ Soc.\ Lond. A {\bf 314} 529-48 (1970).
  
\bibitem{Penroseccc} 
  R. Penrose, Riv.\ Nuovo Cim.  {\bf 1} 252-76 (1969).
  
\bibitem{wald2}
R. M. Wald, Annals of Physics \textbf{82} 548-56 (1974).

\bibitem{hubeny} 
  V.E Hubeny
  Phys.\ Rev.\ D  {\bf 59}, 064013 (1999) [arXiv:gr-qc/9808043].
  
\bibitem{jacobson}
  T.~Jacobson and T.~P.~Sotiriou, Phys.\ Rev.\ Lett. {\bf 103} 141101 (2009) [arXiv:0907.4146v2].
  
\bibitem{barau}
  E.~Barausse, V.~Cardoso and G.~Khanna, Phys.\ Rev.\ Lett. {\bf 105} 261102 (2010) [arXiv:1008.5159v2].

\bibitem{zimmerman} 
  P.~Zimmerman, I.~Vega, E.~Poisson and R.~Haas, Phys.\ Rev.\ D {\bf 87} no. 4 041501 (2010) [arXiv:1211.3889v2]
  
\bibitem{Gao}
  S.~Gao and Y.~Zhang, Phys.\ Rev.\ D {\bf 87} no.4 044028 (2013) [arXiv:1211.2631].

\bibitem{Colleoni} 
  M.~Colleoni and L.~Barack, Phys.\ Rev.\ D {\bf 91} 104024 (2015) [arXiv:1501.07330].
  
\bibitem{hod} 
  S.~Hod, Phys.\ Rev.\ D {\bf 66}, 024016 (2002) [arXiv:gr-qc/0205005].

\bibitem{waldsorce} 
  J.~Sorce and R.~M.~Wald, Phys.\ Rev.\ D {\bf 96} no. 10 104014 (2017) [arXiv:1707.05862].    

\bibitem{toolkit}
  E. Poisson. A relativist's toolkit: the mathematics of black-hole mechanics. Cambridge
university press, 2004.

\bibitem{Natario} 
J.~Natario, L.~Queimada and R.~Vicente, Class.\ Quant.\ Grav.  {\bf 33} no. 17 175002 (2016) [arXiv:1601.06809v2].

  
\bibitem{Chen:2019nhv}
B.~Ning, B.~Chen and F.~Lin, Phys.\ Rev.\ D {\bf{100}} no.4 044043 (2019) [arXiv:1902.00949].

\bibitem{Christodoulou}
D.~Christodoulou, Phys.\ Rev.\ Lett. \textbf{25} 1596-1597 (1970).

\bibitem{bch}
J.~M.~Bardeen, B.~Carter and S.W.~Hawking, Commun.\ Math.\ Phys. \textbf{31}, 161-170 (1973).

\bibitem{will} 
  C.~M.~Will, Astrophys.\ J. {\bf 191}, 521-32 (1974).
  
\bibitem{will2} 
  C.~M.~Will, Astrophys.\ J. {\bf 196}, 41-9 (1975).
  
\bibitem{hod2} 
  S.~Hod, Eur.\ Phys.\ J.\ C {\bf 75}, no. 11, 541 (2015).
  

\bibitem{Duztas:2017ycz}
  K.~Düztaş, Turk.\ J.\ Phys.\  {\bf 42} no.3,  329 (2018).

\bibitem{lohiya} 
  D.~Lohiya, J.\ Phys.\ A {\bf 15}, 1815-21 (1982).

\bibitem{linet} 
  B.~Linet, J.\ Phys.\ A {\bf 9}, 1081 (1976).
  
  
\bibitem{Hod3}
S.~Hod,
Phys. Rev. D \textbf{87} (2013) no.2, 024037

\bibitem{Hod4}
S.~Hod,
Phys. Rev. Lett. \textbf{100} (2008), 121101

\bibitem{needham} 
  T.~Needham, Phys.\ Rev.\ D {\bf 22}, 791 (1980).
  
\bibitem{macke} 
  K.~Mackewicz \& R.~M.~Wald, Phys.\ Rev.\ D {\bf 100}, no. 10, 104043 (2019) [arXiv:1909.03970v3].
  

\end{thebibliography}
\end{document}